\documentclass[%
 reprint,
 amsmath,amssymb,
 aps,
]{revtex4-2}
\usepackage{graphicx}
\usepackage{dcolumn}
\usepackage{bm}
\usepackage{color}

\begin{document}


\title{
Stochastic Gravitational Wave Background from Highly-Eccentric Stellar-Mass Binaries in the Milli-hertz Band
}

\author{Zeyuan Xuan}
\email{zeyuan.xuan@physics.ucla.edu}
\affiliation{ Department of Physics and Astronomy, UCLA, Los Angeles, CA 90095}
\affiliation{Mani L. Bhaumik Institute for Theoretical Physics, Department of Physics and Astronomy, UCLA, Los Angeles, CA 90095, USA}

\author{Smadar Naoz}
\affiliation{ Department of Physics and Astronomy, UCLA, Los Angeles, CA 90095}
\affiliation{Mani L. Bhaumik Institute for Theoretical Physics, Department of Physics and Astronomy, UCLA, Los Angeles, CA 90095, USA}

\author{Bence Kocsis}
\affiliation{Rudolf Peierls Centre for Theoretical Physics, Parks Road, Oxford OX1 3PU, UK}

 \author{Erez Michaely}
\affiliation{ Department of Physics and Astronomy, UCLA, Los Angeles, CA 90095 \\ Mani L. Bhaumik Institute for Theoretical Physics, Department of Physics and Astronomy, UCLA, Los Angeles, CA 90095, USA}

\date{\today}

\begin{abstract}
Many gravitational wave (GW) sources are expected to have non-negligible eccentricity in the millihertz band. These highly eccentric compact object binaries may commonly serve as a progenitor stage of GW mergers, particularly in dynamical channels where environmental perturbations bring a binary with large initial orbital separation into a close pericenter passage, leading to efficient GW emission and a final merger. This work examines the stochastic GW background from highly eccentric ($e\gtrsim 0.9$), stellar-mass sources in the mHz band. Our findings suggest that these binaries can contribute a substantial GW power spectrum, potentially exceeding the LISA instrumental noise at $\sim 3-7$~mHz. This stochastic background is likely to be dominated by eccentric sources within the Milky Way, thus introducing anisotropy and time dependence in LISA's detection. However, given efficient search strategies to identify GW transients from highly eccentric binaries, the unresolvable noise level can be substantially lower, approaching $\sim 2$ orders of magnitude below the LISA noise curve. Therefore, we highlight the importance of characterizing stellar-mass GW sources with extreme eccentricity, especially their transient GW signals in the millihertz band.
\end{abstract}

\maketitle

\section{Introduction}
\label{section:intro}

The study of GW signals from eccentric compact object binaries plays an important role in GW astronomy. For example, extensive efforts have been made to measure the residual eccentricity of GW mergers in the data analysis of LIGO, Virgo, KARAGA (LVK) collaboration, which could shed light on the formation mechanisms of compact binaries \citep[][]{east13, samsing14, Coughlin_2015, Gondan_2018a, Gondan_2018b, moore19,2021ApJ...913L...7A, 2021arXiv211103634T,Zevin_2021,Lower18, Romero_Shaw_2019}. However, so far, no confident evidence of residual eccentricity has been detected \citep[see, e.g.,][]{Abbott_2019ecc,Lenon_2020, Romero-Shaw_2020,gayathri2022eccentricity,Samsing+2022}, primarily because GW radiation tends to circularize the orbit, rendering eccentricity negligible within the LVK frequency band \citep{Peters64,Hiner08}. In the future, the Laser Interferometer Space Antenna (LISA) \citep{2017arXiv170200786A} is expected to observe sources in a lower frequency band ($10^{-4}-10^{-1}~\rm Hz$). Consequently, numerous eccentric GW sources, potentially in their earlier evolutionary stages, may be present in the LISA data stream, providing valuable insights into their surrounding environments \citep[see, e.g., ][]{barack04,mikoczi2012,robson18, chen19, Hoang+19, Fang19, tamanini19, Torres-Orjuela21, amaro+22,Xuan+21, Xuan23}.

Many eccentric GW sources formed through dynamical channels undergo a progenitor stage before the final merger. In this phase, the compact object binary, initially characterized by a large semi-major axis (e.g., $a\gtrsim 0.1 \,\rm au$), attains extreme eccentricity (e.g., $e\gtrsim 0.9$) due to environmental perturbations. Although these sources have an orbital frequency well below the millihertz band, the binary's pericenter distance, $r_p=a(1-e)$, can become sufficiently close to induce strong millihertz GW emission \citep[see, e.g.,][]{Loutrel_2017,Loutrel+20,Xuan+23b}. This process leads to orbital energy loss, causing the orbit to shrink and circularize, resulting in a GW merger.

The GW signal emitted by wide, highly eccentric compact binaries exhibits distinctive characteristics compared to quasi-circular ones. In particular, when the binary's eccentricity is small, the GW signal can be effectively approximated by a near-monochromatic, sinusoidal wave, where the dominant GW frequency is twice the orbital frequency \citep[see, e.g., ][]{Cutler+94}. However, with an increase in the source's eccentricity, the GW emission becomes stronger upon each pericenter passage, turning the signal into a burst-like waveform \citep[e.g., see Fig.1 in reference][]{Xuan+23b}. For example, a compact object binary with eccentricity $e=0.9,0.95,0.99$ will emit more than $88.6\%,99.5\%,99.9996\%$ of its GW energy in less than $1\%$ of the orbital period during the pericenter passage, regardless of its components' masses $m_1,m_2$, and semi-major axis $a$ \citep{Xuan+23b}.

Various dynamic environments can give rise to stellar-mass bursting GW sources. For example, in dense star clusters, compact object binaries may attain non-negligible eccentricity due to external perturbations like GW capture, binary-single, and binary-binary scattering \citep[e.g.,][]{O'Leary+09,Thompson+11,Aarseth+12,Kocsis_2012,breivik16,Gondan_2018a,Orazio+18,Zevin_2019,Samsing+19,Martinez+20,Antonini+19,Kremer_2020,wintergranic2023binary,Gondan_Kocsis2021,purohit2024binary}. Highly eccentric mergers can also arise in stellar disks or active galactic nucleus accretion disks  \citep{Tagawa+2021,Samsing+2022,Munoz+22,Gautham+23}. Additionally, fly-by interactions and galactic tides may excite the eccentricity of wide compact object binaries in the galactic field  \citep[e.g.,][]{Michaely+19,Michaely+20,Michaely+22}, potentially leading to observable mergers. Moreover, in a hierarchical triple system (a tight binary orbiting a third body on a much wider "outer orbit"), the inner binary can undergo eccentricity oscillations via the eccentric Kozai-Lidov (EKL) mechanism \citep{Kozai1962,Lidov1962,Naoz16}, becoming a GW source with high eccentricity in the LISA band. These channels can significantly contribute to the overall merger rate of stellar-mass compact objects \citep[e.g.,][]{wen03,Hoang+18,Hamers+18,Stephan+19,Zevin_2019,Bub+20,Deme+20,Wang+21,Zevin_2021}.

In this paper, our focus is on the stochastic GW background (GWB) originating from these bursting sources. In particular, as a collective GW signal of unresolved sources, the astrophysical stochastic GWB will compete with other expected LISA science sources, potentially producing a confusion noise that affects LISA sensitivity. Previous efforts have aimed to characterize various stochastic backgrounds, such as the collective GW signal from the galactic population of double white dwarf (DWD) binaries \citep[see, e.g., ][]{Nelemans+04,Nissanke+12,Klein+16,lamberts18}, extragalactic sources \citep{barack04,Bonetti+2020,Enoki2007, Huerta2015,Kelley2017,chen17pta}, and cosmological stochastic backgrounds \citep{Barack+04a,Adams+14,Nishizawa_2016,Sesana2016PhRvL,Orazio+18,Boileau+21,kou2024stochastic}. Additionally, it was recently suggested that extreme mass ratio inspirals (EMRIs) might have orders of magnitude higher formation rate \citep{Naoz+22} that may yield an order of magnitude noise level above LISA’s sensitivity level \citep{Naoz+23}. By understanding these backgrounds, we not only address the potential impact of confusion noise on the parameter estimation of other resolved sources but also extract information about the stochastic background itself, thus constraining the population of GW sources \citep[see, e.g., ][]{Agazie_2023,Naoz+23}.

As a natural consequence of dynamical formation, the highly eccentric, stellar mass binaries may create a significant GW background in the LISA band. For example, in our previous studies \citep{Xuan+23b}, we anticipate the presence of approximately 3 to 45 detectable bursting binary black holes (BBHs) within the Milky Way, each with a signal-to-noise ratio (SNR) exceeding 5 for the upcoming LISA mission \citep{Xuan+23b}. Beyond these detectable cases, there could be a considerably larger number of bursting sources with SNR values falling below the detection threshold, yet their collective contribution remains significant. 

Furthermore, the transient nature of these sources poses challenges in extracting their signals from the detector's output \citep{Tai_2014,Loutrel_2017,Loutrel+20}, which potentially leads to a higher noise background level. In particular, the burst detection methods for stellar mass binaries, especially when multiple sources' bursts are present in the data stream simultaneously, remain relatively underdeveloped. Thus, there is uncertainty regarding our ability to identify all bursting sources in the detector's output, even if the sources' SNR is above the detection threshold. Moreover, the astrophysical burst signals may be intertwined with instrumental noise, such as glitches \citep[see, e.g., ][]{spadaro2023glitch,Armano+16,Armano+18}, and contribute to the unresolved GW background. 

We note that there have been previous works examining the GWB from circular stellar mass BBHs \citep[see, e.g., ][]{Nishizawa_2016,Sesana2016PhRvL,Babak2023JCAP...08..034B} and eccentric stellar mass BBHs at a cosmological distance \citep[see, e.g., ][]{chen17,Orazio+18,Zhao2021MNRAS.500.1421Z}. These studies suggested that stellar-mass BBHs can make a non-negligible contribution to future observations, with eccentricity potentially affecting the overall shape of the millihertz GW background. In this work, however, we extend this analysis beyond the cosmological population of BBHs to explore the GWB contribution of highly eccentric sources at close distances, specifically bursting sources in the Milky Way and nearby galaxies. As discussed before, even considering only the Milky Way population, the number of these sources may be substantial \citep{Xuan+23b}, and their close distance can potentially yield a significant overall GWB level. Furthermore, the transient nature of their GW emission can affect the pattern of the GWB, with a single bursting source capable of emitting GWs in a wide range of frequencies, which may cause confusion in future observations. In our analysis, we consider the realistic spatial distribution and non-equilibrium formation history of highly eccentric sources.

We organize this paper as follows. In Section~\ref{sec:burstingproperty}, we first demonstrate the properties of bursting GW sources, then discuss their detectability (Section~\ref{subsec:identification}) and GW background calculation (Section~\ref{subsec:bkgcal_details}). In Section~\ref{sec:simulation}, we introduce the population model of bursting sources, then carry out numerical simulations to predict the GW background level from highly-eccentric, stellar-mass BBHs in the Universe (Section~\ref{subsec:results}). We show the results separately, for the globular clusters (Section~\ref{sec:gcs}), galactic field (Section~\ref{sec:field}),  and the galactic nucleus (Section~\ref{sec:gn}). In Section~\ref{sec:discussion}, we summarize and discuss the overall effect of stochastic GW background from stellar-mass bursting sources.

Unless otherwise specified, we set $G=c=1$.

\section{Theoretical consideration}
\label{sec:theory}
\subsection{The Properties of GW Bursts from Highly Eccentric Binaries}\label{sec:burstingproperty}


In this section, for completion, we briefly summarize our previous results \citep{Xuan+23b} on the detection of GW signals from individual bursting sources. 
The GW emission from a highly eccentric binary is largely suppressed for most of the orbital period, $T_{\rm orb}$. However, it becomes significant during the pericenter passage time, $T_p$, resulting in a GW burst \footnote{Note that we omit an order unity factor of $(1+e)^{-1/2}$, following our previous work, \citep{Xuan+23b}.}\citep[][]{O'Leary+09,Xuan+23b},
\begin{equation}
   T_{p} \sim \frac{r_p}{v_p} \sim (1-e)^{3/2} T_{\rm orb} \ ,
    \label{eq:time0}
\end{equation}
where $v_p$ is the orbital velocity at pericenter, and $T_{\rm orb}=2\pi a^{3/2} m_{\rm bin}^{-1/2}$ is the period of a binary with a mass $m_{\rm bin}$. 

The strain amplitude, $h_{\rm burst}$, and peak frequency, $f_{\rm burst}$ of a single GW pulse in the waveform of a highly eccentric compact object binary, can be estimated analytically \footnote{The peak GW frequency of eccentric sources is often estimated as $f_{\rm peak} = f_{\rm orb} (1+e)^{1/2} (1-e)^{-3/2}$ \citep{O'Leary+09}. A more detailed expression of the peak frequency can also be found in Refs.~\citep{wen03,Randall_2022}. For consistency with other definitions in our previous treatment \citep{Xuan+23b}, we adopt $f_{\rm burst}\sim 2 f_{\rm orb} (1-e)^{-3/2}$, which differs by an order of unity.} (for a detailed explanation, see \citep{Xuan+23b}):
\begin{align}
    f_{\rm burst}&\sim 2f_{\rm orb}(1-e)^{-\frac{3}{2}} \nonumber\\
    &\sim 3.16 {\rm mHz}\, \left(\frac{m_{\rm bin}}{20\rm M_{\odot}}\right)^{\frac{1}{2}} \left(\frac{a}{1\rm au}\right)^{-\frac{3}{2}}\left(\frac{1-e}{0.002}\right)^{-\frac{3}{2}} \ ,
    \label{eq:fpeak}
\end{align}
and
\begin{align}
    h_{\rm burst} &\sim \sqrt{\frac{32}{5}}\frac{m_1m_2}{r_da(1-e)}\nonumber\\
    &\sim  7.65\times 10^{-21}\eta_s \left(\frac{m_{\rm bin}}{20\rm M_{\odot}}\right)^{\frac{5}{3}} \left(\frac{f_{\rm burst}}{3.16\rm mHz}\right)^{\frac{2}{3}}\left(\frac{r_{d}}{8\rm kpc}\right)^{-1}\ ,
        \label{eq:hburst}
\end{align}
in which $f_{\rm orb}=1/T_{\rm orb}$ is the orbital frequency of the bursting source, $r_{d}$ is the comoving distance of the binary, and $\eta_s=4 m_1m_2/(m_1+m_2)^{2}$ is unity for equal mass sources.

Providing the analytical expression of $h_{\rm burst}$ and $f_{\rm burst}$, We can also estimate the signal-to-noise ratio of the bursting source \citep{Xuan+23b}:
\begin{align}
{\rm SNR} \sim
\left\{
\begin{array}{cc}
     \dfrac{h_{\rm burst}}{\sqrt{S_n\left(f_{\rm burst }\right)}} \sqrt{T_{\rm o b s}\left(1-e\right)^{3 / 2}} & (T_{\rm obs} \geq T_{\rm orb})\\[3ex]
     \dfrac{h_{\rm burst}}{\sqrt{f_{\rm burst} S_n(f_{\rm burst })}} & (T_{\rm obs} \leq T_{\rm orb})
\end{array}
\right.
\label{eq:SNR}
\end{align}
here $T_{\rm obs}$ is the observational time of the GW detector, and $S_{\mathrm{n}}(f)$ is the
spectral noise density of LISA evaluated at GW frequency $f$
\citep[see e.g.,][]{2016PhRvD..93b4003K,Robson+19}

In the case of $T_{\rm obs}>T_{\rm orb}$, more than one GW bursts will be detected, and Equation~(\ref{eq:SNR}) can also be expressed as:
\begin{align}
{\rm SNR} \sim &58 \eta_s \left(\frac{m_{\rm bin}}{20\rm M_{\odot}}\right)^{2} \left(\frac{a}{1\rm au}\right)^{-1}\left(\frac{1-e}{0.002}\right)^{-\frac{1}{4}} \nonumber\\
&\left(\frac{r_d}{8\rm kpc}\right)^{-1}\left(\frac{T_{\rm obs}}{4\rm yr}\right)^{\frac{1}{2}} \left(\frac{S_{n}(f_{\rm burst})}{S_{n}(3.16\rm mHz)}\right)^{-\frac{1}{2}}.
\label{eq:SNRnew}
\end{align}

A binary needs to emit GWs in the millihertz band to be potentially detectable with LISA. This condition constrains the orbital parameters of the observed GW sources. For example, a circular binary has its  GW frequency equal twice the orbital frequency, and thus, the orbital radius must shrink to $r_{\rm circ}\sim 10^{-3}\rm au$ (for stellar-mass sources) to yield a GW signal detectable within the millihertz band.
However, for a highly eccentric source, the peak frequency of GW bursts, $f_{\rm burst} \propto f_{\rm orb}(1-e)^{-\frac{3}{2}}$, can reach the millihertz band even when the binary is considerably wide and the orbital frequency is very low \citep{Xuan+23b}. During each orbit, the bursting source only emits GWs for a short period near the pericenter passage, resulting in a much slower orbital energy loss compared to circular binaries with the same GW frequency.

Due to the slower loss of orbital energy, highly eccentric sources could have a significantly longer detectable time within the LISA band. In particular, we can estimate the lifetime of a bursting source, $\tau_{\rm burst}$, by considering the merger timescale of binaries with extreme eccentricity \citep{Peters64,Xuan+23b}:  
\begin{align}
    &\tau_{\mathrm{burst}}\sim \frac{3}{85\mu m_{\rm bin}^2} a^4\left(1-e^2\right)^{7 / 2}\nonumber\\
    &\sim  1.17\times10^{6}{\rm yr} \,\eta_s^{-1}\left(\frac{m_{\rm bin}}{20\rm M_{\odot}}\right)^{-3} \left(\frac{a}{1\rm au}\right)^{4}\left(\frac{1-e}{0.002}\right)^{\frac{7}{2}} ,
    \label{eq:lifetime}
\end{align}
where $\mu=m_{1}m_{2}/(m_1+m_2)$, and $q=m_1/m_2$. Note that this timescale is still much shorter than the Hubble time.

As shown above, the lifetime of millihertz bursting sources is promising, and this timescale is much longer than a millihertz circular binary's merger timescale, providing that circular binary has the same orbital radius, $r_{\rm circ}$, as the bursting sources' pericenter distance $a(1-e)$ \citep{Peters64}:
\begin{align}
\tau_{\rm circ, inspiral} &\sim \frac{5}{256\mu m_{\rm bin}^2}r_{\rm circ}^4 \nonumber\\
&\sim  2.5\times10^{3} {\rm yr} \,\eta_s^{-1} \left(\frac{m_{\rm bin}}{20 \rm M_{\odot}}\right)^{-5/3}\left(\frac{f_{\rm circ, GW}}{3.16\rm mHz}\right)^{-\frac{8}{3}}
\ ,
\label{eq:inspiral}
\end{align}
here $f_{\rm circ,GW}$ is the GW frequency of the circular binary, which equals the eccentric source's burst frequency, $f_{\rm burst}$, in our comparison.

We note that Equation~(\ref{eq:inspiral}) can also be used to estimate the bursting binary's remaining inspiral timescale, when its orbit shrinks and circularizes after the bursting stage, if we replace $r_{\rm circ}$ in the equation with $a(1-e)$, and $f_{\rm circ, GW}$ with $f_{\rm burst}$  \citep[note that the pericenter distance almost keeps constant during the evolution of a highly eccentric GW merger, see][]{Peters64,Xuan+23b}. Therefore, combining Equation~(\ref{eq:lifetime}) and (\ref{eq:inspiral}), we can estimate the ratio between the lifetime of a bursting source and its corresponding moderately-eccentric inspiral stage within the same GW frequency band:
\begin{equation}\label{eq:RBtime}
    \tau_{\rm burst}\sim 20  (1-e)^{-\frac{1}{2}} \tau_{\rm circ, inspiral}\ .
\end{equation}

As mentioned above, dynamically formed GW sources are more likely to have wide eccentric configurations. This stage lasts much longer than the subsequent inspiral with moderate eccentricity, as indicated by Eq.~(\ref{eq:RBtime}). 
Consequently, along the evolution track of a dynamically formed eccentric source, the system will spend a considerably longer time in the first stage (millihertz bursting sources) compared to the second stage (millihertz near-circular inspiral). Therefore, if the dynamical formation channel significantly contributes to the GW merger rate, the potential number of millihertz bursting sources can be much greater than that of circular ones.

We note that Equation~(\ref{eq:RBtime}) is based on the comparison in the same GW frequency band and does not take into account the sources' detectability. However, since circular inspirals have a larger SNR than the highly eccentric configuration, they may exceed the LISA detection threshold below the millihertz band, extending their detectable time and enlarging the detected population. For example, in our previous work \citep{Xuan+23b}, we estimated the detectability of a $10-10{\rm M_{\odot}}$ BBH system with $e\sim0, a\sim 0.03$~au in the Milky Way ($\sim 10\rm kpc$). It turns out that such a system will be detectable for $10^7$~yr, with an orbital period of $T_{\rm orb}\sim 10^{4}$~s and GW frequency $f_{\rm circ,GW}\sim 10^{-4}$~Hz. Because we expect a large population of circular DWDs in the sub-millihertz band \citep[see, e.g.,][]{Nissanke12, lamberts18}, the identification of the near-circular, low-frequency systems is beyond the scope of this study. Thus, in this paper, we focus on the dynamically formed, highly-eccentric bursting sources and limit our aforementioned argument of the population in the same frequency band (e.g., $f_{\rm GW}\gtrsim 1\rm mHz$).

\subsection{Stochastic GW Background from Bursting Sources}
\label{sec:bkgcalculation}
\subsubsection{Identification of Bursting Signals and Residual Noise}
\label{subsec:identification}
\begin{figure}[htbp]
    \centering
    \includegraphics[width=3.5in]{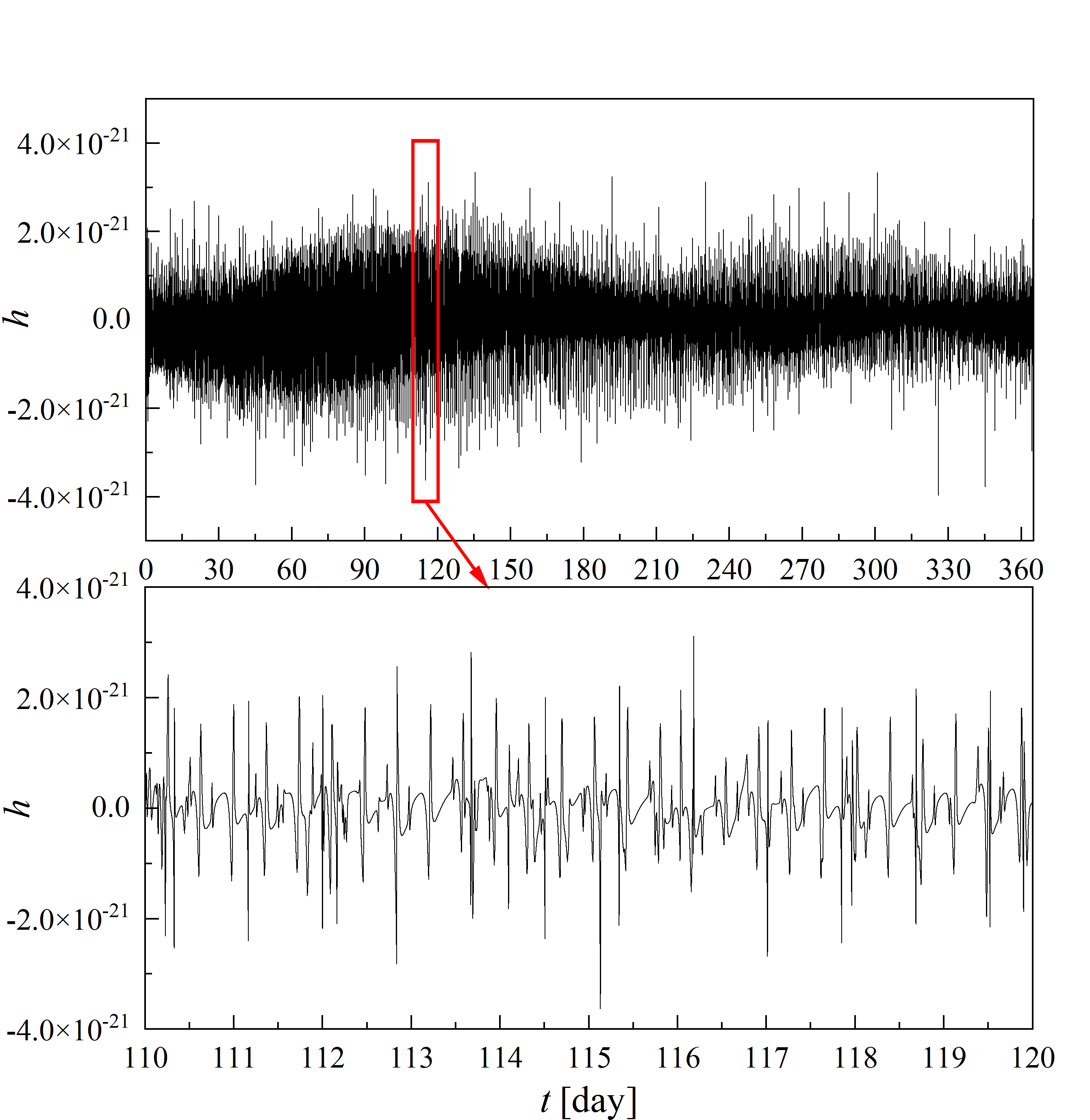} 
    \caption{{\bf{GW background from bursting BBHs within the Milky Way globular clusters.}} Here we generate the bursting BBHs population using the methodology described in Section~\ref{sec:simulation}, and calculate their collective GW signals observed by LISA. {\it Upper panel} shows the detector's Michelson response \citep{Cutler+98,Cornish+03}, $s_1(t)$, over a 1-yr period. {\it Lower panel} shows a zoom-in of the signal. In the {\it upper panel}, the modulation of the GW background is a result of the detector's annual motion around the sun.
    }
    \label{fig:gcbkgeg}
\end{figure}

Bursting sources have the potential to form a non-negligible GW background in the millihertz band, given their significant GW radiation (Equations~(\ref{eq:fpeak})-(\ref{eq:SNRnew})) and extended lifetime (Equation~(\ref{eq:lifetime})). Particularly, frequent bursts from multiple sources with low SNR can accumulate to yield a stochastic background noise (see Eqs.~(\ref{eq:SNR}) and (\ref{eq:SNRnew})). For example, in Figure \ref{fig:gcbkgeg}, we show the simulation result of the collective GW signal from bursting BBHs in the Milky Way globular clusters. We randomly generate BBH systems and estimate their ages based on a constant formation rate $\gamma_{\rm GC}=10^{-6}\,\mathrm{yr}^{-1}$ in the Milky Way (for details of the population model, see Section~\ref{sec:simulation}). At the time of observation, we numerically compute the GW signal of each individual system (excluding merged ones) using the x-model \citep{Hinder+10}, then combine them to obtain the collective GW background in the time domain. Additionally, we also include the LISA detector's response to the bursting GW background, which yields a long-term modulation due to the orbit of LISA around the Sun (see {\it Upper Panel} of Figure \ref{fig:gcbkgeg}). As shown in this example, the collective bursting signal is a superposition of GW transients, i.e., individual bursts, from various sources, each localized in time and frequency.

The transient nature of bursting signals (see {\it Lower Panel} of Figure \ref{fig:gcbkgeg}) poses several challenges for existing data analysis strategies, such as the matched filtering method \citep{Finn:1992xs, Cutler+94}, which relies on accurate templates of GW signals and may not perform optimally in fitting bursts \citep[see, e.g.,][]{Tai_2014, Loutrel_2017, Loutrel+20}. In particular, GW templates for highly eccentric sources are underdeveloped, which adds to the difficulty of constructing a template bank for the extraction of the signal. Moreover, multiple bursting sources can be present in the detector's output, which may lead to the misidentification of sources and degeneracy of the fitted parameters.

We note that many efforts have been made to analyze transient events in LIGO data analysis (e.g., power stacking \citep{east13}, wavelet decomposition \citep{Klimenko_2004}, and the Q-transform \citep{bassetti2005development,Tai+14}). Furthermore, several studies focused on 
the transients from highly eccentric EMRIs, which can be possibly seen by LISA  \citep{barack04,Cornish+03b,porter2010eccentric,mikoczi2012,Fan+22}. However, the burst detection methods for stellar mass binaries, particularly when multiple sources' bursts are present in the datastream simultaneously, remain relatively underdeveloped. In other words, there is uncertainty regarding our ability to identify all bursting sources, even when their signal-to-noise ratio is sufficiently high.

Therefore, we divide the following discussions into two parts, with different criteria for bursting source identification, to demonstrate the potential noise level in future data analysis: 
\begin{enumerate}
    \item {\bf Total GW background} This criterion includes all the bursting sources regardless of their $\rm SNR$, assuming that we do not include burst transient templates in the future LISA data analysis. It yields an upper limit for the potential stochastic background level from these sources.
    \item {\bf Unresolvable noise background} This criterion only includes the bursting sources with overall $\rm SNR<8$ for a $4$~yr LISA mission, assuming that we successfully identify all the bursting sources with overall $\rm SNR >8$ and distinguish them from other signals. It gives a conservative estimation of the noise level from bursting sources.
\end{enumerate}

\subsection{Stochastic Background Calculation}\label{subsec:bkgcal_details}
\subsubsection{Stochastic Background from GW Sources with Fixed Orbital Parameters}
We follow \citet{Barack+04a} to calculate the stochastic GW background caused by stellar-mass bursting binaries in the LISA band, in which the one-sided noise spectral density is estimated as:
\begin{equation}
    S_n(f)=\frac{4}{\pi f^2} \frac{d \rho}{d f} ,
    \label{eq:snf0}
\end{equation}
here $f$ is the observed frequency, and $d\rho/df$ is the total GW energy density in the corresponding frequency band.

For a single eccentric binary, its time domain waveform, $h(a, e, t)$, can be decomposed into different harmonics, each with the frequency of $f_n$ \citep[for simplicity, here we rms averaged over the
binary inclination, see e.g., ][]{peters63,oleary09,Kocsis_2012,Kocsis2012b}:
\begin{equation}
h(a, e, t)=\sum_{n=1}^{\infty} h_{\mathrm{c},n}\left(a, e\right) \exp \left(2 \pi i f_n t\right),
\label{eq:hecc}
\end{equation}
where
\begin{equation}\label{eq:hcn}
f_n=nf_{\rm orb}\,,\quad
   h_{\mathrm{c},n} = 
   \frac{2}{n} \sqrt{g(n, e)} h_0(a)
\end{equation}
and
\begin{equation}
    h_0(a)=\sqrt{\frac{32}{5}} \frac{m_1 m_2}{r_d a} \ ,
    \label{eq:h0}
\end{equation}
\begin{align}
g(n, e)= & \frac{n^4}{32}\left[\left(J_{n-2}-2 e J_{n-1}+\frac{2}{n} J_n+2 e J_{n+1}-J_{n+1}\right)^2\right. \nonumber\\
& \left.+\left(1-e^2\right)\left(J_{n-2}-2 J_n+J_{n+2}\right)^2+\frac{4}{3 n^2} J_n^2\right],
\label{eq:g}
\end{align}
in which $J_i$
is the i-th Bessel function evaluated at $ne$. Note that $r_{d}$ is the comoving radial distance to the binary. 

The energy density of a monochromatic GW signal with frequency $f$ can be expressed as \citep[see, e.g.,][]{Maggiore_2000}:
\begin{equation}
    \rho_{\rm GW}=
    \frac{\pi}{8} f^2\left\langle h_{+}^2+h_{\times}^2\right\rangle \sim \frac{\pi}{4}f^{2}h_0^{2},
    \label{eq:rhocirc}
\end{equation}
where $h_{+},h_{\times}$ are the amplitude of GW's two polarizations, and $h_0$ stands for the sky-averaged stain amplitude in Equation~(\ref{eq:h0}). 

Since the gravitational wave signal from an eccentric source consists of multiple harmonics with different frequencies, the energy density is a superposition of each harmonic's contribution. In other words, by combining Equations~(\ref{eq:hecc})-(\ref{eq:rhocirc}), we can sum the contributions of all harmonics to determine the total gravitational wave energy density of an eccentric binary system:
\begin{equation}
\rho_{\rm ecc}=\sum_{n=1}^{\infty} \frac{\pi}{4} f_n^2 h_{\mathrm{c},n}^2 \ .
\label{eq:rhoecc1}
\end{equation}

Here, we assume the GW source evolves slowly. Thus, the signal does not undergo a significant frequency shift during the observation, and each harmonic is a monochromatic signal where the power concentrates on $f=nf_{\rm orb}$ (see Equation~(\ref{eq:rhocirc})). We expect that most sources will be consistent with this assumption for the harmonics that contribute most to the detection signal-to-noise ratio \citep[e.g.,][]{O'Leary+09,Wang+21,Xuan+23b}; however, in some cases, the signal may shift in a short timescale, \citep{Hoang+19,Deme+20}, which is beyond the scope of this study.  

A key term in the noise spectral density, Eq.~(\ref{eq:snf0}), is $d\rho/df$. To get this term from Equation~(\ref{eq:rhoecc1}), we need to calculate the distribution of energy density as the function of the GW signal's frequency. In other words, we sum the energy density contributions from the harmonics within a given frequency bin $(f_i,f_f)$:
\begin{equation}
\Delta \rho_{\rm ecc}\bigg|_{f = f_i}^{f = f_f}  =\sum_{n=n_i}^{n_f} \frac{\pi}{4} f_n^2
h_{\mathrm{c},n}^2 \approx \int_{n_i}^{n_f}\frac{\pi}{4} f_n^2
h_{\mathrm{c},n}^2 dn\ .
\label{eq:rhoecc2}
\end{equation}
Here, $n=n_i,n_i+1, ...\,n_f$ represents all the harmonic numbers between the initial and final frequencies, $f_i/f_{\rm orb}$ and $f_f/f_{\rm orb}$. 
Furthermore, as can be seen from Equation~(\ref{eq:hecc})-(\ref{eq:rhoecc2}), the number of harmonics making a substantial contribution to the signal rises rapidly as eccentricity reaches extreme values. For example, we may have to consider millions to billions of harmonics when calculating the spectral density of GW signal from a binary with $e\gtrsim0.95$. Consequently, using the sum in Equation~(\ref{eq:rhoecc2}) might not always be practical in noise spectral density calculations. 
Thus, for the highly eccentric systems in this paper, we average Equation (\ref{eq:rhoecc2}) over the frequency bin to get a smoothed expression of $\rho(f)$.
In particular, consider a frequency bin with the width of $df$ at frequency $f$. Each GW harmonic in this bin contributes an energy density of approximately $\sim \frac{\pi}{4} f^2 h_{\mathrm{c},n(f)}^2$, where $n= f/f_{\rm orb}$, and the number of harmonics included in this bin equals $dn = df/f_{\rm orb}=n df/f$. Therefore, the energy density in this bin for a single source, $d \rho_{\rm{ecc, single}}$, is obtained by multiplying the energy density of a single harmonic by the number of harmonics:
\begin{equation}
    d \rho_{\rm{ecc, single}}(f)=\frac{\pi}{4} f^2 h_{\mathrm{c},n(f)}^2 \times \frac{n(f) d f}{f},
    \label{eq:avgrho1}
\end{equation}

and thus,

\begin{equation}
    \frac{d \rho_{\rm ecc,single}}{d f}=\frac{\pi}{4} f h_{\mathrm{c},n(f)}^2 n(f).
    \label{eq:avgrho2}
\end{equation}

Note that $n$ in Equation~(\ref{eq:avgrho1}) and (\ref{eq:avgrho2}) is a function of $f$.

Below, we substitute Equation~(\ref{eq:avgrho2}) into \eqref{eq:snf0} to calculate the energy spectrum of a single bursting source, then sum all the bursting sources' contributions to get the total GW power spectrum:
\begin{equation}
    S_{n,\rm loc}(f)=\frac{4}{\pi f^2} \frac{d \rho}{d f}=\sum_{i\in\rm sources} \frac{n_i(f)  h_{\mathrm{c},n_i(f)}^2}{f}
    \label{eq:rho}\ ,
\end{equation}
where $n_i(f)=f/f_{{\rm orb},i}$ for the $i^{\rm th}$ source. Equation~(\ref{eq:rho}) is useful when calculating the bursting GW background from a group of local sources with determined orbital parameters (e.g., bursting BBHs in the Milky Way galactic center, at a given time in the numerical simulation).

\subsubsection{Time-Averaged Stochastic Background Under the Steady State Approximation}\label{subsub:avgsnf}
    For some bursting BBH populations, we assume a continuous birth and death of compact object binaries while keeping the total number of GW sources unchanged (i.e., adopt the steady-state approximation). Therefore, to obtain the expectation value of total GW power spectral density, we further average Equation~(\ref{eq:rho}) over the evolution of highly eccentric systems. 
    
    In particular, first, consider the contribution of the $i^{\rm th}$ bursting source in Equation~(\ref{eq:rho}). The average of this contribution over the single system's time evolution is: 
    \begin{equation}
    \label{eq:rho_avgtime}
    \langle S_{n,\rm loc}(f)\rangle_{\rm single, i^{\rm th}} = \int_{t_0}^{t_0+\tau_{\rm merger}}
    \frac{  h_{\mathrm{c},n_i(f)}^2(f,t)}{f_{{\rm orb}, i}(t)}\frac{dt}{\tau_{\rm merger}} \ ,
    \end{equation}in which $h_{\mathrm{c},n}(f, t) = 
   \frac{2}{n(f,t)} \sqrt{g(n(f,t), e(t))} h_0(a(t))$ (see Equation~(\ref{eq:hcn})), and $n(f, t)= f/f_{\rm orb}(t)$. 
   
   In Equation~(\ref{eq:rho_avgtime}), $t_0$ is the formation time of the BBH system in the simulations, and $\tau_{\rm merger}$ represents the merger time of the BBH system with initial orbital parameters $a(t_0),e(t_0)$. When considering the long-term evolution of the source, parameters such as orbital frequency ($f_{\rm orb}$), semi-major axis ($a$), and eccentricity ($e$) become functions of time. For bursting systems undergoing isolated evolution, these parameters can be determined using Eqs.~5.6 - 5.14 in \citet{Peters64}. However, for bursting systems affected by dynamical effects, such as the eccentric Kozai-Lidov mechanism, numerical simulations are required to determine the merger time and the evolution of other orbital parameters.
    
    When integrating Equation~(\ref{eq:rho_avgtime}) over the lifetime of a GW source, we determine its time-averaged GW power spectral density. Under the steady-state approximation, where the ensemble average equals the time average, Equation~(\ref{eq:rho_avgtime}) provides the expectation value of spectrum contribution from the $i^{\rm th}$ GW source in observation.
    
    Furthermore, we average Equation~(\ref{eq:rho_avgtime}) over different BBH systems' initial conditions in the population synthesis. This average is computed by summing the GW power spectral density of all the simulated systems, $\langle S_{n,\rm loc}(f)\rangle_{\rm single, i^{\rm th}}$ (see Equation~(\ref{eq:rho_avgtime})), and dividing by the number of systems generated in the simulation, $N_{s}$:
    
    \begin{equation}
    \langle S_{n,\rm loc}(f)\rangle\bigg|_{\rm single, avg}= \frac{1}{N_{s}}
   \sum_{\rm i=1}^{N_{s}}\langle S_{n,\rm loc}(f)\rangle_{\rm single, i^{\rm th}} \ .
    \label{eq:rho_avgtime2}
    \end{equation}
    We note that $N_{s}$ is artificially introduced to enlarge the sample size and ensure numerical accuracy. It differs from the actual number of systems observed. 
    
    Finally, we multiply Equation~(\ref{eq:rho_avgtime2}) with the expected number of bursting BBHs in observation, $\langle N_{\rm tot} \rangle$, and get the collective GW power spectrum:
        
    \begin{equation}
    \langle S_{n,\rm loc}(f)\rangle\bigg|_{\rm tot}= \frac{\langle N_{\rm tot} \rangle}{N_{s}}
   \sum_{\rm i=1}^{N_{s}}\langle S_n(f)\rangle_{\rm single,i} \ .
    \label{eq:rho_avgtime3}
    \end{equation}
    Here, $\langle N_{\rm tot}\rangle=\langle \tau_{\rm merger}\rangle \gamma_{\rm merger}$, where $\langle \tau_{\rm merger}\rangle$ represents the average lifetime of merger systems in the simulation, and $\gamma_{\rm merger}=dN/dt$ denotes the BBH merger rate per unit time in the considered region (see Section~\ref{sec:configuration}). Equation~(\ref{eq:rho_avgtime3}) is useful when calculating the expectation of bursting GW background from local sources in the Monte Carlo simulation (e.g., bursting BBHs in the Milky Way globular clusters, assuming they have reached equilibrium distribution with a fixed merger rate.).

\subsubsection{Average Stochastic Background from Extragalactic Sources}
For bursting sources at a cosmological distance, we further take into account the redshift $z$ \citep[e.g., see][]{Naoz+23,Barack+04a}:
\begin{equation}
    S_{n,\rm cos}(f)=\frac{4}{\pi f^2} \int_0^z \frac{d t}{d z} \dot{\varepsilon}(f(1+z),z) d z,
    \label{eq:noisecos}
\end{equation}
here $\dot{\varepsilon}(f_{\mathrm{em}},z)\Delta f_{\mathrm{em}} $ represents the rate (per unit proper time, per unit co-moving volume) at which GW energy between frequencies $f_{\mathrm{em}}$ and $f_{\mathrm{em}} + \Delta f_{\mathrm{em}}$ is emitted into the universe, at red-shift z. $f_{\mathrm{em}}=f(1+z)$ denotes the emitted GW frequency (i.e., the frequency measured by a contemporaneous observer, not the redshifted frequency $f$ we measure
today). We note that Equation~(\ref{eq:noisecos}) has taken into account the decrease in the waves’ energy due to the redshift. For a detailed explanation, see Eqs.~34 - 35 in \citet{Barack+04a}. 

Furthermore, we calculate $\dot{\varepsilon}(f_{\mathrm{em}},z)$ using the following relation:
\begin{equation}
    \dot{\varepsilon}(f_{\mathrm{em}},z)= \Gamma(z) \left\langle \frac {dE_{\mathrm{GW}}\left( f_{\mathrm{em}}\right)}{df_{\mathrm{em}}}\right\rangle.
    \label{eq:noisecos2}
\end{equation}

In Equation~(\ref{eq:noisecos2}), $\Gamma(z)=dN/(dtdV)$ represents the comoving merger rate of bursting BBHs per unit proper time, per unit co-moving volume; $\left\langle dE_{\mathrm{GW}}\left( f_{\mathrm{em}}\right)/df_{\mathrm{em}}\right\rangle$ is the emitted GW energy per unit frequency from a bursting BBH system, averaged over different BBH systems' initial conditions from the population synthesis.

We note that, the meaning of $dE_{\rm GW}/df_{\mathrm{em}}$ in Equation~(\ref{eq:noisecos2}) is different from $d\rho_{\rm single}/df_{\mathrm{em}}$ in Equation~(\ref{eq:avgrho2}). In particular, $E_{\rm GW}$ accounts for the total gravitational wave energy emitted throughout the evolution of a source, while $\rho_{\rm single}$ represents the source's gravitational wave energy density at a given location in space (or equivalently, the energy flux $F_{\rm single}=c\rho_{\rm single}$, where $c=1$ is the speed of light). These two quantities can be related by integrating the energy flux $F$ on a spherical surface $\Omega_s$, with a fixed radius of $d_l$ from the BBH system, throughout the sources' time evolution. (see, e.g., Eq.~25 in \citet{phinney2001practical})

\begin{equation}
\int_{t_0}^{t_0+\tau_{\rm merger}}\langle F_{\rm single}(f_{\rm em},t)\rangle_{\Omega_s} d t = \frac{1}{4 \pi d_l^2} \int_0^{\infty}\frac{d E_{\rm GW}}{d f_{\rm em}} d f_{\rm em},\label{eq:fluxandenergy}
\end{equation}
and thus:
\begin{align}
\frac{d E_{\rm GW}}{d f_{\rm em}} &=4 \pi d_l^2\frac{d}{d f_{\rm em}} \int_{t_0}^{t_0+\tau_{\rm merger}}\langle F_{\rm single}(f_{\rm em},t)\rangle_{\Omega_s} d t \nonumber \\ &=4 \pi d_l^2 \int_{t_0}^{t_0+\tau_{\rm merger}}\left\langle \frac{d\rho_{\rm single}(f_{\rm em},t)}{d f_{\rm em}}\right\rangle_{\Omega_s} d t,\label{eq:fluxandenergy2}
\end{align}

In Equation~(\ref{eq:fluxandenergy2}), $F_{\rm single}=c\rho_{\rm single}$ can be expressed as a function of $f_{\rm em}$ and $t$. This is because, at any given time during the system's evolution, $\rho_{\rm single}$ is a function of the emitted GW frequency (see Equation~(\ref{eq:avgrho2})). However, over the long-term evolution, other orbital parameters also become functions of time (see the explanation below Equation~(\ref{eq:rho_avgtime})). Therefore, when integrating over the time evolution of a GW source, Equation~(\ref{eq:fluxandenergy2}) yields the cumulative energy emitted within a fixed frequency bin.

We note that, Eq.~25 in \citet{phinney2001practical} includes the effect of redshift at arbitrarily large luminosity distance, while here we drop the term of $1+z$ in Equation~(\ref{eq:fluxandenergy}). This is because our purpose is to get the total energy measured as a function of the emitted frequency in the comoving frame of the source (the cosmological effects have been taken into account via Equation~(\ref{eq:noisecos})). In other words, we choose the spherical surface $\Omega_s$ with a short distance $d_l$ from the BBH system, such that the effect of redshift is negligible when we sum over the energy flux. In fact, the factor $d_l^2$ will cancel out with the amplitude term in $\rho_{\rm single}$ on the right side of Equation~(\ref{eq:fluxandenergy2}), thus its value will not affect the calculation of $d E_{\rm GW}/d f_{\rm em}$, since the energy density $\rho_{\rm single}$ is proportional to the square of the strain amplitude (see Equation~(\ref{eq:avgrho2})), while the strain amplitude is inversely proportional to the distance (see Equations~(\ref{eq:hcn}) - (\ref{eq:h0})).

We emphasize that, in general, there exists a straightforward relationship between the spectrum of the gravitational wave background produced by a cosmological distribution of discrete gravitational wave sources and
the present-day comoving number density of remnants, as described by Eq. 5 in Ref.~\citep{phinney2001practical}. Realistic examples and further discussions on this relationship can be found in studies such as Refs.~\citep{Kocsis2011gasmerger,Kocsis2012b}.  For our cases, however, since the local population can have a non-negligible contribution to the overall GW background level, we include the inhomogeneous spatial distribution and non-equilibrium star formation history in the Monte Carlo simulations detailed below (see Section~\ref{sec:configuration}). Therefore, for consistency, we adopt Equations~(\ref{eq:rho}), (\ref{eq:rho_avgtime3}) and (\ref{eq:noisecos}) as a robust approach to estimate the GW background level across all channels, including cosmological background calculations.

\section{Simulations of Different Bursting Galactic Sources }
\label{sec:simulation}
\subsection{Configuration of the Simulations}
\label{sec:configuration}
As a proof of concept, we adopt the population model in \citet{Xuan+23b} to generate the parameters of highly eccentric, stellar-mass BBHs in the millihertz band \footnote{Note that for other kinds of bursting compact object binary, such as Double White Dwarfs (DWDs), there exists considerable uncertainty in their orbital evolution due to tidal interaction and mass transfer \citep[see, e.g.,][]{kremer18,Lau_2022}, particularly for cases involving extreme eccentricity. Therefore, our discussion is limited to BBHs, and we leave the discussion of other populations for future work. Nevertheless, it's worth noting that the tidal interaction of DWDs tends to circularize the orbit, preventing eccentricity from reaching extreme values \citep[see, e.g.,][]{Shara_2002, Willems_2007}. Additionally, the close approach of bursting DWDs at their pericenter may induce mass transfer and spin interaction, further restricting the lifetime of highly eccentric sources \citep[see, e.g., Fig.4 in][]{Xuan+23b}. Thus, we suspect a limited population of highly eccentric DWDs and a negligible noise background from bursting DWDs in the millihertz band, even in the presence of external dynamical perturbations.}. We consider three regimes that are expected to host eccentric BBHs: Globular Clusters (GCs), Galactic Field, and Galactic Nuclei (GNs). Within each of these regimes, we split the discussion into several specific cases specified below. 

In the simulation of BBHs in Globular Clusters and the Galactic Field, we adopt the steady-state approximation, assuming a continuous birth and death of compact object binaries in the universe while keeping the total number of GW sources unchanged. We calibrate our GW power spectrum based on the BBHs merger rate in the universe: $\Gamma_{\rm GC}=7.2_{-5.5}^{+21.5}\, \mathrm{Gpc}^{-3}\, \mathrm{yr}^{-1}$ \citep[e.g.,][]{Rodriguez2016, Kremer_2020, Antonini_2020}, $\Gamma_{\rm Field, Fly-by}=5_{-2}^{+5}\, \mathrm{Gpc}^{-3}\, \mathrm{yr}^{-1}$ \citep{Michaely+19}. Moreover, adopting the galaxy number density of $0.02~\rm Mpc^{-3}$ \citep{Conselice_2005}, we can use the BBH merger rate in the universe to estimate the merger rate in an averaged galaxy: $\gamma_{\rm GC}=3.6_{-2.7}^{+10.8}\times 10^{-7}\,\mathrm{yr}^{-1}$, $\gamma_{\rm Field, Fly-by}=2.5_{-1}^{+2.5}\times 10^{-7}\,\mathrm{yr}^{-1}$.

For the Galactic Nuclei, we consider not only the old population of stars (calibrated using the steady-state approximation and the $m-\sigma$ relation \citep{merritt1999black, Kormendy+13}), but also incorporate the non-equilibrium star formation history in the case of the Milky Way center, based on observational results \citep[see, e.g., ][]{Paumard06, Lu_2008, Do2013, Chen+23}.

We note that the merger rate of BBHs can depend on the star formation history and exceed the estimation in the local universe at high redshift. This phenomenon is especially significant for the globular cluster channel \citep[see, e.g., ][]{PortegiesZwart_2000,Gnedin_2014,Rodriguez2015,Rodriguez2016gcmerger,Fragione2018,Fragione2018MN}. Therefore, when calculating the cosmological background of bursting GW sources from GCs, we additionally take into account the change of BBH merger rate as a function of redshift and integrate Equation~(\ref{eq:noisecos}) up to redshift $z=3$ (i.e., $11.5$Gyr from present), following the merger rate evolution in figure 1 of Ref.~\citep{Fragione2018}. The adopted redshift cutoff is partly justified because $z=3$ represents the epoch of the peak of globular
cluster formation rate \citep{Gnedin_2014}.

For a more detailed description of the population model, see Sections 3.2-3.6 and Appendix A of Ref.~\citep{Xuan+23b}.

\subsection{Stochastic background from the Simulations' Results}
\label{subsec:results}

\subsubsection{ Dynamically Formed BBHs in Globular Clusters}
\label{sec:gcs}
    
    Globular clusters are proposed to be a primary channel for the formation of GW mergers. In particular, the dense stellar environment within globular clusters results in various dynamical interactions, including scattering, few-body captures, the Eccentric Kozai-Lidov Mechanism, and non-secular triple interactions \citep[see, e.g.,][]{banerjee10,Orazio+18, Samsing+18, Antonini+19, Fragione+19, Martinez+20, Kremer_2020}. Due to these frequent dynamical interactions, BBHs in globular clusters are anticipated to have non-negligible eccentricity, even within the frequency band of LIGO \citep[see, e.g.,][]{Orazio+18, Antonini+19, Zevin_2019, Samsing+19, Martinez+20, Kremer_2020}. Moreover, since globular clusters widely exist in the universe (e.g., approximately 150 in the Milky Way \citep{Harris+96, Baumgardt+18}), we expect a significant number of bursting sources originating from GCs.

    For the GW background calculation, we adopt the BBH eccentricity and semi-major axis distribution from previous studies \citep[see, e.g., figure 4 in][for a summary]{Martinez+20}, and assume that all the BBHs have the mass of $10-10~\rm M_{\odot}$ for simplicity. Since we are interested in the bursting stage where BBHs are in a wide configuration, we evolve the systems shown in \citet{Martinez+20}, which is in a higher frequency band, back to the former bursting stage. With the aforementioned parameter distribution, we carry out Monte Carlo simulations, randomly generate BBH systems, and evolve them until the merger happens. The stochastic background is calculated under the steady-state approximation, using Equation~(\ref{eq:rho_avgtime3}).

    We adopt the spatial distribution of GCs in the Milky Way from Ref.~\citep{Arakelyan_2018}.  For simplicity, we assume that the detector's distance to all GCs in M31 (M33) is 777 (835) kpc, respectively. The cosmological background of bursting BBHs in GCs is determined using Equation~(\ref{eq:noisecos}), with the assumption that the number density of Milky-Way like galaxies is $0.02~\rm Mpc^{-3}$ \citep{Conselice_2005}.

    \begin{figure*}[htbp]
    \centering
    \includegraphics[width=3.45in]{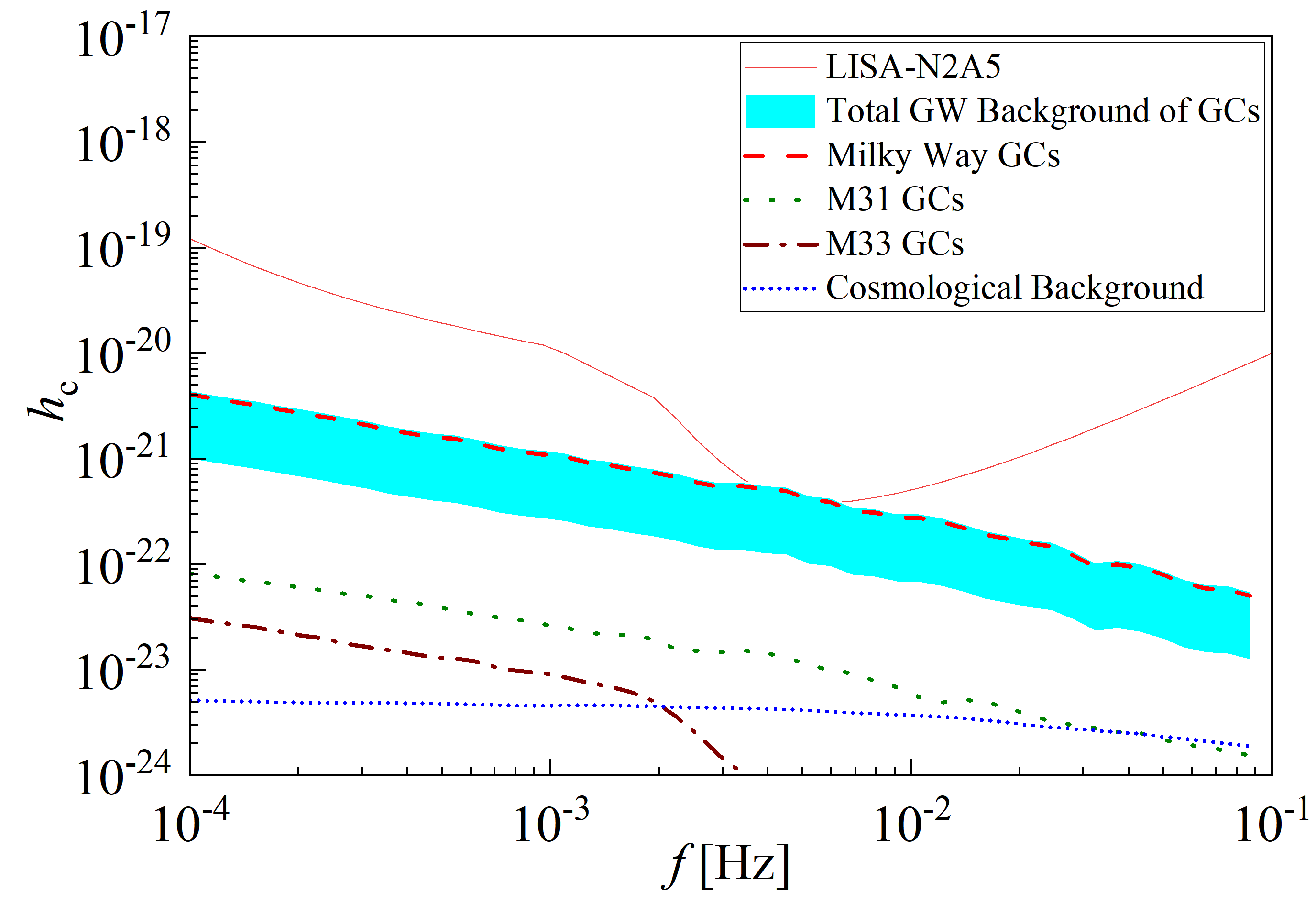} 
    \includegraphics[width=3.55in]{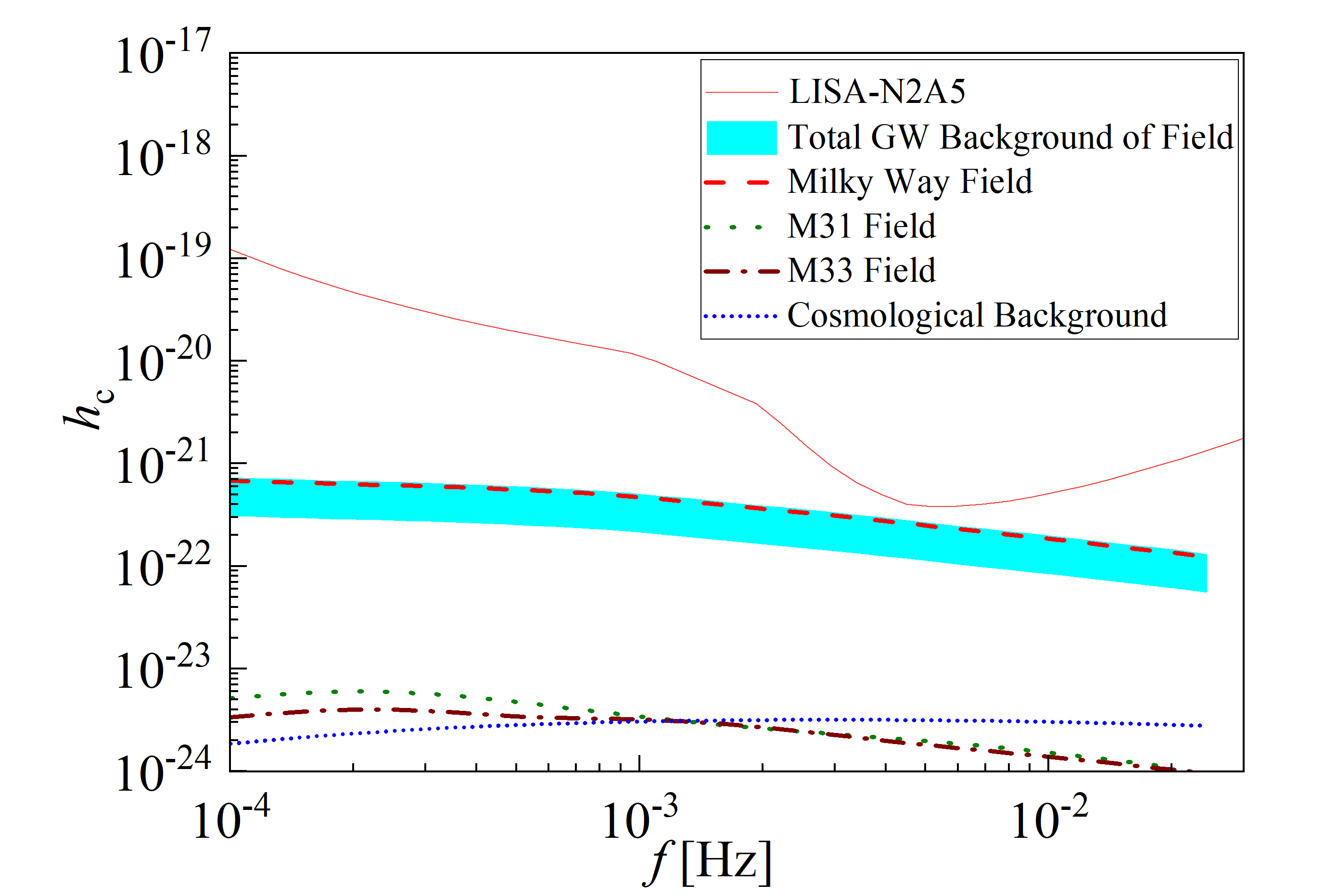} 
    \caption{{\bf Total GW background from bursting BBHs in Globular Clusters and Galactic Field.} Here we show  (in dashed lines) the maximum collective GW power spectrum, $\sqrt{fS_n(f)}$, corresponding to bursting BBHs in the GCs ({\it Left Panel}) and Galactic Field ({\it Right Panel}) of Milky Way, M31, M33, and cosmological sources. The blue-colored region shows the total GW background from all four channels mentioned above, with its boundary representing the upper and lower limits. For comparison purposes, we also plot the LISA noise curve (N2A5 configuration \citep{2016PhRvD..93b4003K,Robson+19, 2024lisa}, in the solid red line).
    }
    \label{fig:gcs}
    \label{fig:field}
    \end{figure*}

Figure \ref{fig:gcs}, {\it Left Panel}, shows the maximum stochastic background from bursting BBHs in the Milky Way GCs (red dashed line), M31 GCs (green dotted line), M33 GCs (brown dashed-dot line), and GCs at a cosmological distance (blue dotted line). The overall GW power spectrum (depicted by the blue-shaded band) is obtained by summing the contributions of all the sources mentioned above, where the maximum (minimum) value corresponds to the maximum (minimum) aforementioned merger rate. As can be seen in  {\it Left Panel} of Figure~\ref{fig:gcs}, bursting BBHs within Milky Way GCs dominate the overall GW power spectrum. Meanwhile, the contributions from the other three channels have comparable levels, approximately two orders of magnitude weaker than Milky Way GCs. Moreover, the total GW background reaches the LISA noise curve at around 5 mHz, which suggests that neglecting these sources in future data analysis could significantly compromise LISA's sensitivity.

We note that previous studies \citep[see, e.g., ][]{Sesana2016PhRvL,Nishizawa_2016,Orazio+18} have shown the overall cosmological GW background from stellar-mass BBHs, accounting for all the eccentricity values, is either comparable to or smaller than the millihertz LISA noise level. For our cases, as the highly eccentric binaries represent a subset of the entire BBH population, we anticipate their cosmological GW spectrum level to be lower than the overall background from BBHs and, thus, fall below the LISA noise curve. This prediction serves as an upper bound for the cosmological background of bursting sources in the simulation and turns out to be consistent with the simulation result (depicted by the blue dotted line in Figure~\ref{fig:gcs}).

Additionally, the noise background from the unresolvable sources (i.e., the GW spectrum of sources with $\rm SNR$ below 8 for a 4-year LISA mission), is over two orders of magnitude weaker than the LISA sensitivity. Thus, it is omitted here to avoid clutter, and we present this below, see Figure \ref{fig:summarynoise}.
    

    \subsubsection{Fly-by Induced, Highly Eccentric BBHs in the Galactic Field} 
    \label{sec:field}
    
    Recent studies show that external perturbations, such as flybys or galactic tides, can exert a significant influence on the orbital evolution of wide compact object binaries in the galactic field \citep[see, e.g.,][]{kaib14,Michaely+19, Michaely+20, Michaely+22}. Although the isolated wide binaries are not likely to merge due to their large GW merger timescale (sometimes exceeding the Hubble time, as indicated by Equation (\ref{eq:inspiral})), the dynamical perturbations from the environment can drive their eccentricity to extreme values, making the BBH system emit GW bursts during pericenter passage, and resulting in a GW merger.  Given the substantial number of stars in the galactic field, the fly-by-induced bursting BBHs can significantly contribute to the overall GW background, even if their fraction within the entire stellar population is small.

    We follow a similar approach for the GW background calculation as shown in Section~\ref{sec:gcs}. In particular, we generate the spacial distribution of wide BBHs according to the density profile of a Milky-Way type galaxy (see Eqs.~23 - 26 in Ref.~\citep{Michaely+19}). We randomly generate the semi-major axis of BBHs with the log-uniform distribution (from $100$ to $50000~\rm au$), then calculate the fly-by merger rate as a function of semi-major axis \citep[see Eqs.~16 - 23 in][]{Michaely+22}. With a given semi-major axis, there is a critical eccentricity, $e_{\rm crit}$, for a BBH system to merge before the next fly-by \citep[see Eq.~2 in][]{Michaely+22}. Therefore, we randomly generate the initial eccentricity of fly-by induced merger in the range of $e_{\rm crit}$ to $1$, following the thermal distribution $F(e)=2e$. 
    
    Following the abovementioned approach, we compute the parameter distribution of fly-by-induced bursting BBHs, then adopt Equation~(\ref{eq:rho_avgtime3}) to estimate the overall GW background of Milky Way, M31, and M33, under the steady state approximation. For the cosmological background, we fix the BBHs merger rate $\Gamma_{\rm Field, Fly-by}=5_{-2}^{+5}\, \mathrm{Gpc}^{-3}\, \mathrm{yr}^{-1}$ for simplicity, and integrate Equation~(\ref{eq:noisecos}) up to $z=3$ to keep consistent with the case of GCs (see Section \ref{sec:configuration}). As shown in the {\it Right Panel} of Figure~\ref{fig:field}, although the total GW background from bursting BBHs in the galactic field does not reach the LISA noise curve level, it still significantly contributes to the overall GW background.

    \subsubsection{ BBHs in the Galactic Nuclei}
    \label{sec:gn}
    The galactic nucleus, particularly if hosting an SMBH, can be a promising environment for the formation of stellar-mass bursting BBHs \citep[see, e.g.,][]{Kocsis_2012,Rodriguez2016, Hoang+18, Hoang_2019, Trani19,Stephan+19, Wang+21, Arca+23,Zhang24}. For example, consider a stellar-mass BBH system orbiting the SMBH in the galactic nucleus. The EKL mechanism can induce eccentricity excitation in the inner binary's orbit, potentially driving the eccentricity to extreme values and triggering GW bursts \citep{wen03,Naoz16,Hoang+18,Hamers+18,Stephan+19,Bub+20,Deme+20,Wang+21}. Consequently, we anticipate a high fraction of bursting BBHs in the galactic center, especially within the Nuclear Star Cluster, where the influence of the SMBH dominates the dynamical environment.

    The number of bursting BBHs in the Galactic Nucleus can strongly depend on their formation history. For example, the main population of stars in the Milky Way Galactic Nuclei formed approximately $2\sim 8~\rm\, Gyr$ ago \citep{Lu_2008,Chen+23}. Consequently, this old population of stars may have reached an equilibrium state, resulting in a low rate of compact object binary formation. Therefore, we can use the steady-state approximation and adopt the same approach outlined in Section~\ref{sec:gcs} and \ref{sec:field} to calculate their GW power spectrum. 
    
    \begin{figure}[htbp]
    \centering
    \includegraphics[width=3.5in]{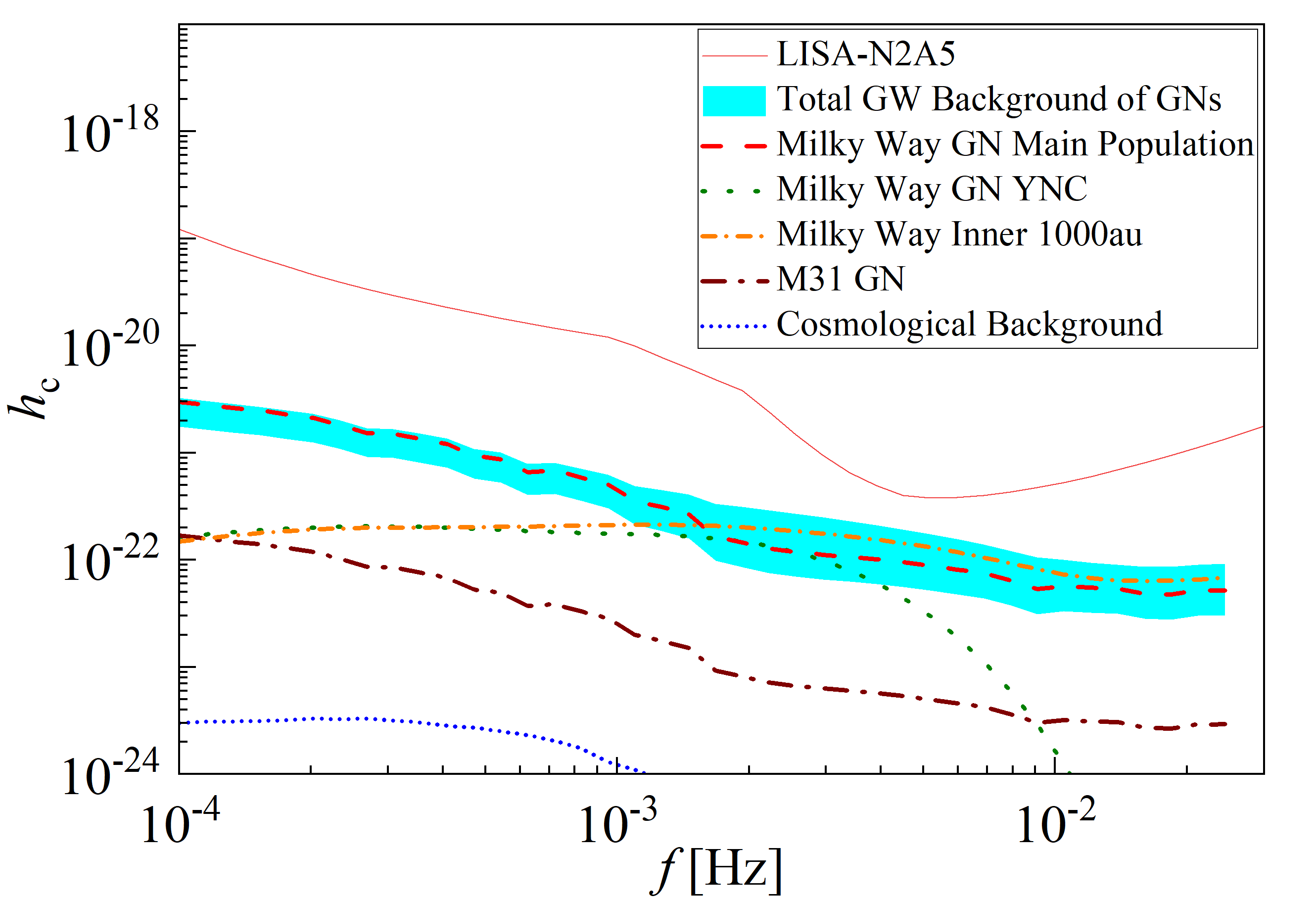}
    \caption{{\bf Total GW background from bursting BBHs in the Galactic Nuclei.} Here we show the same GW background results as in Figure~\ref{fig:gcs}, but for the case of bursting BBHs in the galactic nuclei. We exclude the case of M33 in this figure since it does not have a SMBH in the galactic center. For a detailed explanation of the populations shown in the figure, see Section~\ref{sec:gn}.
    }
    \label{fig:gn}
    \end{figure}

    However, observation shows there is a young nuclear star cluster (YNC) at the Milky Way center, which is formed approximately $2\sim 8~\rm\, Myr$ ago \citep[see, e.g., ][]{Paumard06, Lu2006, Do2013, Chen+23}. Moreover, there might be a hidden mass of $\sim 3000~\rm M_{\odot}$ within the inner $1000$~au of the Milky Way center \citep[e.g.,][]{Do+19,Gravity+20}, supported by the observation of stellar motion and the theoretical arguments of the stability of S0-2, \citep[e.g.,][]{Chu18,chu+23,Naoz+20,Zhang+23,Will+23}. In our previous work \citep{Xuan+23b}, we carried out Monte-Carlo simulations of these BBHs' evolution, taking into account the EKL effect up to the octupole level of approximation \citep{Naoz+13}, general relativity precession \citep[e.g.,][]{naoz13}, and GW emission \citep{Peters64}. The results indicated that the young population of stars potentially has a much higher fraction of bursting BBHs, especially when the YNC's age is below the detectable lifetime of bursting sources \citep[see, e.g., Fig.7 in][]{Xuan+23b}. 

    The summary of the simulation results is presented in Figure~\ref{fig:gn}. For the main population of stars in the Milky Way center, the galactic nuclei of nearby galaxies (M31), and the cosmological background, we follow a similar approach as mentioned in Sections~\ref{sec:gcs} and \ref{sec:field}, adopt the steady-state approximation, and use the simulation results in \citet{Xuan+23b} to calculate their GW background.
    
    However, the steady-state assumption is no longer suitable for the bursting BBHs in the YNC and inner $1000$~au of the Milky Way center. Thus, we track the time evolution of these populations as a function of the time, assuming that they were born in a starburst. The total GW power spectrum is calculated using Equation~(\ref{eq:rho}) and then averaged over the possible age range of the system. The blue-colored region in Figure~\ref{fig:gn} represents the collective GW power spectrum from all these channels' contributions. When computing its upper boundary, we sum over the maximum possible GW background level of all these channels. For its lower boundary, we exclude the contribution from bursting BBHs in the inner $1000$~au, as the existence of this population is highly uncertain.

\section{Discussion}
\label{sec:discussion}




Highly eccentric compact object binaries with wide orbits can naturally arise from the dynamical formation of gravitational wave sources. During each pericenter passage, these systems emit significant gravitational waves, potentially resulting in burst signals within the millihertz band. We adopted the analysis from our previous paper \citep{Xuan+23b} and characterized the gravitational wave burst using Equations (\ref{eq:fpeak}) - (\ref{eq:SNRnew}), detailed in Section \ref{sec:burstingproperty}, for completeness.

While the population of bursting sources may be large in the Universe, primarily due to their extended lifetime (see Equation~(\ref{eq:lifetime})-(\ref{eq:RBtime})), their GW signals exhibit a transient nature, which poses challenges in the data analysis. In other words, current data analysis approaches may not properly identify bursting gravitational wave sources, even when the overall power of a signal (i.e., SNR) exceeds the conventional detection threshold for circular binaries (see the discussion in Section \ref{subsec:identification}). Therefore, highly eccentric binaries can contribute to a non-negligible stochastic GW background in future millihertz detections (as depicted by the example in Figure~\ref{fig:gcbkgeg}).

In Section~\ref{subsec:bkgcal_details}, we developed an analytical framework for computing the stochastic GW background from bursting sources. Subsequently, we carried out numerical simulations to determine the GW background from bursting BBHs in different formation channels, utilizing the population model of highly eccentric binaries outlined in Section~\ref{sec:configuration}. Specifically, Figure \ref{fig:gcs} depicts the example of the GW stochastic background from GCs and the Galactic Fields (left and right panels, respectively), adopting a steady-state approach for the sources' formation rate. In Figure \ref{fig:gn}, we also include the fact that a young population of stars has been observed in the center of the Milky Way \citep{Lu2006,Do2009,Lu2013,Do2013,Do+19,Chen+23}. This young population implies that the center of our galaxy may deviate from the steady state restriction. Thus, we explore the impact of a young population in the nuclear star cluster 
on the stochastic background. It produces a comparable contribution to the GW background (see Figure \ref{fig:gcs}, dotted green line). Additionally, there have been observational and theoretical several studies that suggested the existence of about a few thousand of solar mass inwards to S0-2's orbit \citep{Gravity+20,Naoz+20,Zhang+23,Will+23,Jurado+23}. We thus add a population of compact objects inwards to S0-2's orbit ($\sim 1000$~au). That may also have a significant contribution to the GW stochastic background.





As depicted in Figures~\ref{fig:gcs}-\ref{fig:gn}, highly eccentric BBHs within the Milky Way dominate the stochastic background of GW bursts across all three channels (Globular Cluster, Galactic Field, and Galactic Nuclei), surpassing the contributions from other galaxies and the cosmological background. This pattern indicates that highly eccentric binaries are localized GW sources, like the galactic DWDs.

The dominance of the Milky Way's contribution, or in other words, the reason for the low level of cosmological background, can be understood by comparing the GW energy of bursting and non-bursting sources at the same frequency band (i.e., with the same pericenter distance $r_p=a(1-e)$, see Equation~(\ref{eq:fpeak})). According to steady-state approximation, the number of GW sources is proportional to their lifetime, making the total number of bursting sources approximately $\sim 20(1-e)^{-1/2}$ times the number of circular inspirals formed via dynamical channels (see Equation~(\ref{eq:RBtime})). However, since highly eccentric sources have lower GW emission power, the energy density contribution of a single bursting source is suppressed by a factor of $\sim (1-e)^{3/4}$ in the same frequency band (see Equation~(\ref{eq:SNR})). Consequently, the overall power spectrum of bursting sources is suppressed by a factor of $(1-e)^{1/4}$ compared to the contribution from all near-circular sources, implying that their signals are generally weaker, and detection is primarily limited to a short distance. 

\begin{figure}[htbp]
    \centering
    \includegraphics[width=3.5in]{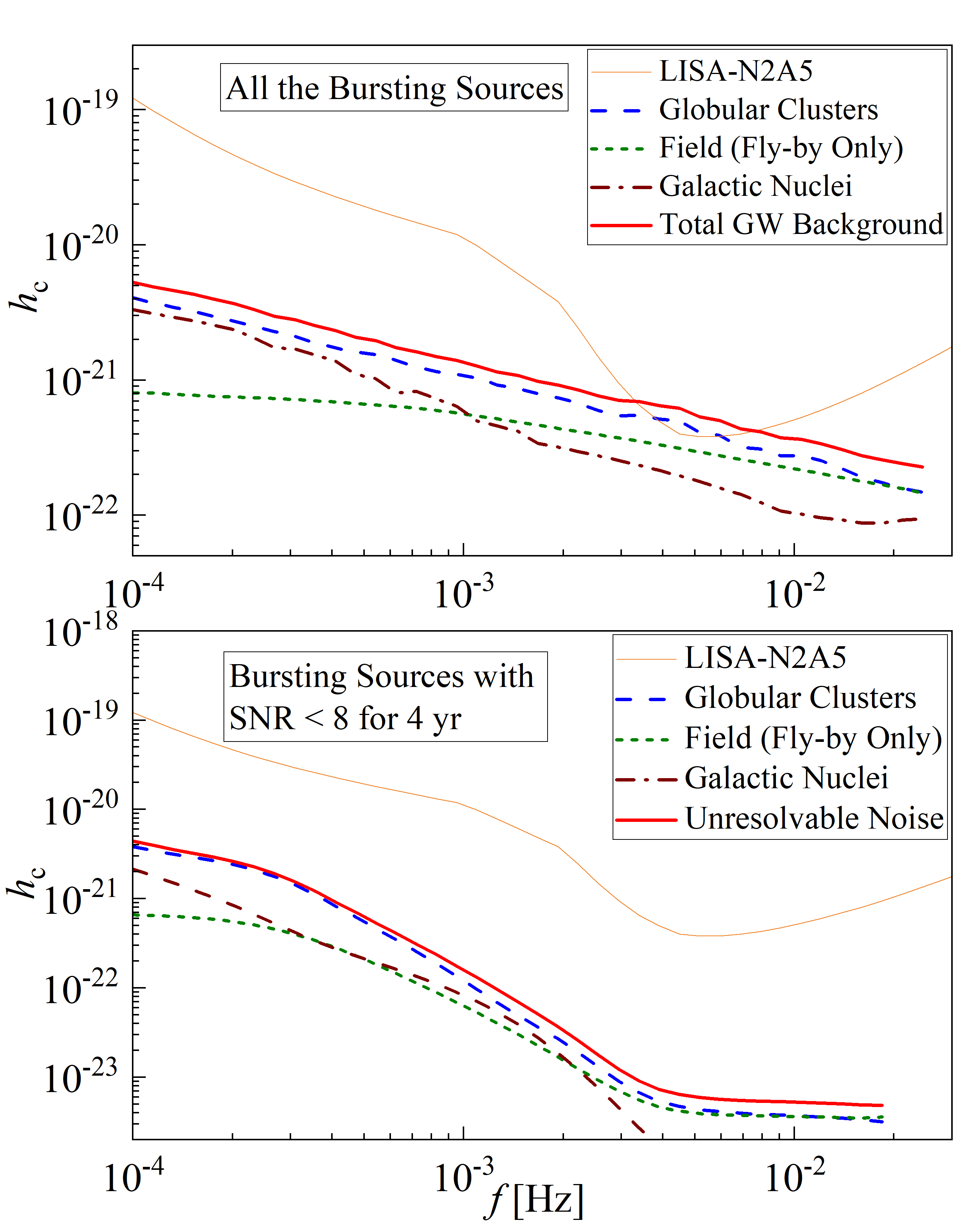}
    \caption{{\bf Summary of the stochastic GW backgrounds from bursting BBHs.} Here, we show the upper limit of the GW background, $\sqrt{fS_n(f)}$, corresponding to highly eccentric BBHs from the Globular Clusters (blue dashed lines), Galactic Field (green dashed lines), and Galactic Nuclei (brown dash-dotted lines) in the universe. {\it Upper Panel} shows the maximum level of total GW background from bursting BBHs. {\it Bottom Panel} represents the unresolvable GW noise background from bursting sources with $\rm SNR<8$ for a $4$~yr LISA mission. For demonstration purposes, we sum the contribution of the three channels mentioned above and show the collective background in the solid red lines, and plot the LISA noise curve (N2A5 configuration \citep{2016PhRvD..93b4003K,Robson+19}, in orange). }
    \label{fig:summarynoise}
\end{figure}

In Figure~\ref{fig:summarynoise}, we summarize the stochastic GW background from bursting BBHs. In particular, {\it Upper Panel} shows the maximum level of total GW background from bursting BBHs in Globular Clusters, the Galactic Field, and the Galactic Nuclei, corresponding to the upper boundary of the blue-colored region in Figures \ref{fig:gcs} and \ref{fig:gn}.
We depict the cumulative contribution of the three channels mentioned above with the solid red line. 
{\it Bottom Panel} represents the unresolvable GW noise background from bursting sources with $\rm SNR<8$ for a $4$~yr LISA mission, which serves as a conservative estimation of the noise level under the assumption that we can identify all the bursting signals with proper templates in the data analysis. 

As can be seen in the figure, highly eccentric, stellar mass BBHs can have a non-negligible GW background contribution in the LISA detection, potentially exceeding the LISA instrumental noise at $\sim 3-7$~mHz (see {\it Upper Panel}, in Figure \ref{fig:summarynoise}). This phenomenon is especially significant if we do not have an effective search strategy for these sources' GW signals. In the LISA data analysis, if we only use the near-circular GW templates, the GW bursts from these sources will not be identified and will only be considered as a noise background 
in the detector's output. Therefore, even with promising $\rm SNR$, the bursting GW sources will become part of the unresolvable noise background, thus affecting the LISA sensitivity (see Section~\ref{subsec:identification}). 

However, providing that all the stellar mass bursting sources are properly identified in the future, the residual noise level is around two orders of magnitude lower than the LISA sensitivity curve, thus having a limited effect on the data analysis (see the {\it Bottom Panel}). Considering the abovementioned points, we highlight the importance of searching and identifying bursting GW sources in LISA detection. Particularly, we may need to develop portable GW templates for these sources in the matched filtering or adopt other time-frequency approaches, such as power stacking \citep{east13} and the Q-transform \citep{bassetti2005development,Tai+14}, to characterize the GW bursts from astrophysical sources in LISA data analysis.

Additionally, in realistic detection, the antenna pattern of LISA undergoes rotation across the sky, causing its response to gravitational waves to change with time. As the bursting BBHs in the Milky Way predominantly contribute to the stochastic background and may have an anisotropic distribution, the amplitude of the GW burst background can vary considerably throughout the year. This observation is consistent with the pattern illustrated in Figure~\ref{fig:gcbkgeg} \citep[see, e.g.,][for a similar case]{Digman_2022}.

We note that the population model used in this study is subject to large uncertainties related to star formation and detailed dynamical evolution. It is intended as a heuristic estimate of the GW background from bursting sources. However, as demonstrated in this paper, the anticipated stochastic background of GW bursts can be significant, even when considering only Milky Way binary black holes. Furthermore, the noise level of GW bursts highly depends on the adopted data analysis strategy in LISA detection. Hence, we emphasize the significance of characterizing stellar-mass GW sources with extreme eccentricity, especially their transient GW signals in the millihertz band.

\acknowledgments
We thank the anonymous referee for useful comments. ZX acknowledges partial support from the Bhaumik Institute for Theoretical Physics summer fellowship. ZX, SN, and EM acknowledge the partial support from NASA ATP 80NSSC20K0505 and from NSF-AST 2206428 grant and thank Howard and Astrid Preston for their generous support. This work was supported by the Science and Technology Facilities Council Grant Number ST/W000903/1 (to BK).
\appendix

\bibliographystyle{apsrev4-1.bst}
\bibliography{bibbase}

\end{document}